\documentclass[a4paper]{jpconf}
\usepackage{graphicx}

\newcommand{\lesssim}{\lower.5ex\hbox{$\; \buildrel < \over\sim \;$}}
\newcommand{\gtrsim}{\lower.5ex\hbox{$\; \buildrel > \over\sim \;$}}
\newcommand{\uu}{{\epsilon_{14}}}

\begin{document}
\title{Best-Bet Astrophysical Neutrino Sources}

\author{Charles D.\ Dermer}

\address{Code 7653, Naval Research Laboratory, 
4555 Overlook Ave., Washington, DC 20375-5352}

\ead{dermer@gamma.nrl.navy.mil}

\begin{abstract}
Likely astrophysical sources of detectable high-energy ($>>$ TeV)
neutrinos are considered. Based on gamma-ray emission properties, the
most probable sources of neutrinos are argued to be GRBs, blazars,
microquasars, and supernova remnants.  Diffuse neutrino sources are
also briefly considered.
\end{abstract}

\section{Introduction}

This discussion of sources of high-energy astrophysical neutrinos
focuses on candidate discrete sources of neutrinos (see also
\cite{lm00}). Arguments for detectable neutrino sources are based
on the $\gamma/\nu$ connection: photohadronic and secondary nuclear
processes, which are the most important astrophysical neutrino
production mechanisms, will also produce $\gamma$ rays.  By
identifying the brightest and most fluent $\gamma$-ray sources, we
therefore identify the most likely neutrino point sources. This
argument is, however, far from airtight.  A source can be $\gamma$-ray
bright without being neutrino bright if the $\gamma$-rays originate
from leptonic processes, for example, when relativistic electrons
Compton scatter ambient target photons to $\gamma$-ray
energies. Conversely, a bright neutrino source can be $\gamma$-ray dim
if the $\gamma$ rays are attenuated.  This can occur either in a
buried source, where the surrounding material or the local radiation
field provides large opacity to $\gamma$ rays or, for cosmologically
distant objects, when high-energy $\gamma$-rays are attenuated by the
extragalactic background light (EBL).

At EGRET and GLAST energies ($E \approx 100$ MeV -- 10 GeV),
attenuation by the EBL is unimportant, even for the highest redshift
objects. Thus the EGRET catalog can be enlisted to identify the
brightest $\gamma$-ray sources and, by the $\gamma$-$\nu$ argument,
the most probable neutrino point sources. The new discoveries with
HESS, MAGIC, and VERITAS at TeV energies provide additional important
information, especially for galactic sources.

Diffuse $\gamma$-ray emissions include those that are genuinely
diffuse, namely cosmogenic GZK neutrinos and neutrinos produced by
cosmic-ray interactions in our Galaxy, as well as apparently diffuse
neutrino fluxes from the superposition of unresolved sources, such as
star-forming galaxies and clusters of galaxies.

\section{The $\gamma$-$\nu$ connection, 
$\gamma$-ray sources, and candidate neutrino sources} 

Photohadronic and secondary nuclear production processes are the two
principal mechanisms responsible for high-energy neutrino
production. The dominant near-threshold channel for neutrino
production through either of these processes involves the excitation
of a $\Delta^+$ isobar. For proton/photon and proton/nucleon
interactions, these reactions take the form
\begin{equation}
p+\gamma^\prime, ~p+ N \rightarrow  
\Delta^+ \rightarrow \cases{ p+\pi^0 \rightarrow p + 2\gamma
&$~$\cr
	 n+\pi^+ \rightarrow n + e +3\nu 
\rightarrow p + 2e+4\nu &$~$\cr}\;,
\label{pgamma}
\end{equation}
where $\gamma^\prime$ represents a target photon and $N$ represents a
target nucleon.  The charge-changing reaction takes place one-third of
the time. For a large number of such reactions, therefore, the outcome
is a high-energy lepton and three high-energy neutrinos for every four
high-energy $\gamma$ rays.  In addition, neutron $\beta$-decay forms
an electron and a neutrino with $\approx 10^2$ times less energy than
the other secondaries.  Approximately four times as much energy is
radiated in electromagnetic secondaries as in neutrinos.  Any
high-energy neutrino source will therefore be a strong $\gamma$-ray
source if the $\gamma$ rays reach the observer without intervening
attenuation.
 
For such $\gamma$-ray sources, the ``best-bet" neutrino sources are
those where a neutrino telescope such as IceCube can detect a neutrino
above the cosmic-ray induced background. IceCube's background is 
at the level $\gtrsim 0.4 \nu(> 1$ TeV)per yr per 
square degree and  $0.08 \nu(> 10$ TeV) per yr per 
square degree \cite{kar03}.  To be detected with a km-scale neutrino
telescope such as IceCube, the neutrino fluence, and therefore the
photon fluence, must be at the level of $\Phi_\gamma\gtrsim 10^{-4}$
ergs cm$^{-2}$, as we now show \cite{da06}.

The detection probability for muon neutrinos is 
$P_{\nu_\mu}(\uu) \approx  10^{-4} \epsilon_{14}\;,
{\rm~for~} 0.1 \lesssim \epsilon_{14} \lesssim 10\;,$
where $\epsilon_{14} =  \epsilon_\nu/100 {\rm~TeV}= 
\epsilon_\nu/160 {\rm~ergs}$ 
\cite{ghs95}. 
Consider a neutrino source with number flux $\phi_\nu(\epsilon_{14}) =
dN/dAdtd\uu = K\epsilon_{14}^{-s}$ in the energy range $0.1 \lesssim
\epsilon_{14} \lesssim 10$. The neutrino energy fluence received
during the time interval $\Delta t$ is $\Phi_\nu ({\rm ergs~cm}^{-2})
\cong (160{\rm~ergs})\Delta t\int_{0.1}^{10} d\uu\;\uu\;\phi_\nu(\uu
). $ The detection of $N_{\nu_\mu}$ muon neutrinos occurs when $N_\nu
\cong A \Delta t \int_{0.1}^{10} d\uu\; P_{\nu_\mu}(\uu
)\phi_\nu(\epsilon_{14}) = 10^{-4} A \;{\Phi_\nu ({\rm ergs~cm}^{-2})/
160{\rm~ergs}} > 1\;, $ where $A$ is the detector area.  Thus the
detecttion of one $\nu_\mu$ with a km-scale neutrino detector ($A
\cong 10^{10}$ cm$^2$) requires that the neutrino fluence $\Phi_\nu
\gtrsim 10^{-4} {\rm ~ergs~cm}^{-2}\;.$ By the $\gamma$-$\nu$
connection, this means that the $\gamma$-ray fluence from a source
must be $\Phi_\gamma \gg 10^{-4}$ ergs cm$^{-2}$.

EGRET reported results of their observations in units of
$10^{-8}\phi_{-8}$ ph($>$100 MeV) cm$^{-2}$ s$^{-1}$ \cite{har99}.
For a flat $\nu F_\nu$ spectrum, the mean photon energy in the EGRET
range, $\approx 100$ MeV -- 5 GeV, is $\approx 400$ MeV, which
corresponds to a bolometric $\nu F_\nu$ flux in this energy range of
$\approx 6\times 10^{-12}\phi_{-8}$ ergs cm$^{-2}$ s$^{-1}$.  The
integral photon flux sensitivity for a two-week on-axis pointing with
EGRET is at the level of $\phi_{-8} \approx 15$, so that the limiting
sensitivity of EGRET for a $5\sigma$ detection was $\approx 10^{-10}$
ergs cm$^{-2}$ s$^{-1}$.  To reach a fluence of $10^{-4}\Phi_{-4}$
ergs cm$^{-2}$ therefore requires that $\Delta t \phi_{-8} \cong
1.5\times 10^7 \Phi_{-4}$ s. The standard EGRET observation lasted for
two weeks. When taking into account Earth occultation, the typical
observing time was $\approx 6\times 10^5$ s. Thus sources with
$\phi_{-8} \gg 30$ fulfill the requirement that if neutrinos are
produced with comparable fluence as $\gamma$ rays, then they would be
detectable neutrino sources. Even though the EGRET energy range is at
a much lower value than that of the neutrino telescopes, the $\gamma$
rays produced in association with the neutrinos could cascade down
into this energy range. Extrapolating the $\gamma$-ray spectrum into
the PeV range means that bright $\gamma$-ray sources with nearly flat
$\nu F_\nu$ spectral indices (or photon number indices $\alpha_{ph}
\approx 2$, defining photon number fluxes $\phi_\gamma(\epsilon)
\propto \epsilon^{-\alpha_{ph}}$) are good candidate neutrino sources.

By examining the Third EGRET catalog \cite{har99}, one finds many
$\gamma$-ray sources that fit this criterion.  The following objects
have at least one and sometimes many two-week observing periods during
which $\phi_{-8} \gtrsim 100$: the blazars PKS 0208-512, PKS 0528+134,
NRAO 530, 3C 279, PKS 1622-297; pulsar 1706-44 and the Crab, Vela and
Geminga pulsars; the EGRET sources associated with the supernova
remnants W44, IC 443, and $\gamma$ Cygni; the sources 3EG J1824-1514
and 3EG J0241+6103 associated with the microquasars LS 5039 and LSI
+61 303, respectively; and several unidentified EGRET sources,
including some in the Galactic plane and some at high latitude. In
addition to these persistent or flaring sources, a bright Solar flare
was detected with EGRET, and several GRBs were strongly detected with
the EGRET spark chamber. We discuss these various classes of sources
in increasing likelihood of being detectable neutrino sources.

\subsection{Solar Flares}

The June 11, 1991 Solar flare radiated a $> 100$ MeV energy fluence
exceeding $10^{-4}$ ergs cm$^{-2}$ \cite{kan93}. The flare spectrum
was fit by a slowly decaying ($\sim 255$ minutes) pion emission
component and a fast-decaying ($\sim 25$ minutes) electron
bremsstrahlung component. The very soft $\gamma$-ray spectrum, with
number index $\gtrsim 3$ -- 4, and lack of evidence for $\gg$ GeV
proton and ion acceleration in Solar flares (for example, from
ground-based neutron monitors) make it unlikely that very high-energy
neutrinos are produced by Solar flares, though they could be sources
of GeV -- TeV neutrinos.
 
\subsection{Pulsars and Pulsar Wind Nebulae}

Pulsars are the brightest point sources in the EGRET catalogs, with $>
100$ MeV $\nu F_\nu$ fluxes of some pulsars exceeding $10^{-9}$ ergs
cm$^{-2}$ s$^{-1}$. Moreover, they are persistently
bright. Consequently it only took EGRET $\approx 1$ day to measure
pulsed fluences $\gtrsim 10^{-4}$ ergs cm$^{-2}$.  It is unlikely,
however, that they would be detectable neutino sources. The pulsed
spectrum always cuts off below several hundred GeV. If this is due to
an electromagnetic cascade, as expected in both polar cap or outer gap
models for pulsed $\gamma$-ray emission, then the emission originates
from electron acceleration and cascades, which would not produce
neutrinos.

The nebulae formed by the cold but highly relativistic MHD winds
expelled by rotating, highly magnetized neutron stars could accelerate
protons and ions that would subsequently undergo interactions to
produce neutrinos. The Crab nebular emission is at the level of
$10^{-10}$ ergs cm$^{-2}$ s$^{-1}$ in the EGRET energy band, but is
convincingly explained as the self-Compton component from the
electrons that radiate the synchrotron nebular emission.

The pulsar wind nebulae discovered at TeV energies with HESS could be
the result of accelerated protons that interact with ambient material
to form the power-law spectra measured from these sources. In its
preliminary galactic plane scan \cite{aha05a}, HESS reached $5\sigma$
sensitivities at the level of $3\times 10^{-11}$ ph$(> 100$ GeV)
cm$^{-2}$ s$^{-1}$ implying, for a mean photon energy of 400 GeV, a
limiting bolometric sensitivity of $2\times 10^{-11}$ ergs cm$^{-2}$
s$^{-1}$. For example, the pulsar wind nebula in the supernova remnant
MSH 15-52 radiated $\approx 3.3
\times 10^{-11}$ ergs cm$^{-2}$ 
s$^{-1}$ in the energy range 0.3 -- 40 TeV with $\alpha_{ph} = 2.27\pm
0.2$ \cite{aha05b}. Extrapolating into the PeV range gives a flux of
$\approx 5\times 10^{-12}$ ergs cm$^{-2}$ s$^{-1}$, or a fluences
$\gtrsim 10^{-4}$ ergs cm$^{-2}$ in $\approx 1$ years time.  Because
of similar morphology to the keV synchrotron emission, the TeV
$\gamma$ rays are likely to be the result of Compton-scattering rather
than hadronic emission.  If this emission is the result of hadronic
processes, however, such sources could be marginally detectable
neutrino sources (see \cite{bbm05,kb06} for a more on Galactic neutrino
sources).

\subsection{Supernova Remnants}

Some of the brightest EGRET sources are associated with supernova
remnants \cite{tor03}, and they display relatively hard spectra
($\approx -2$). They are also believed to accelerate cosmic rays to
the knee of the spectrum ($\approx 3$ PeV).  If due to $\pi^0$ decay
emission from secondary nuclear production, the spectra will soften at
energies well above the $\pi^0$ peak, which is at several hundred MeV
in a $\nu F_\nu$ representation.  In the TeV energy range, a
bolometric $\nu F_\nu$ flux reaching $10^{-10}$ ergs cm$^{-2}$
s$^{-1}$ is observed with HESS \cite{aha06a} from RX J1713.7-3946. One
would optimistically expect \cite{ah02} this SNR to be detected as a
neutrino sources after only some months of observing except,
unfortunately, the $\gamma$-ray spectrum exhibits a remarkable cutoff
above $\approx 10$ TeV, though IceCube should still detect
one or two $\nu_\mu(>1$ TeV)/yr \cite{kb06}. The cutoff might indicate that acceleration to
higher energies has not yet occurred, though the remnant should be
well into the Sedov phase. The higher energy cosmic ray
protons and ions, being more diffusive, could leave the
acceleration region quickly. The emission could also be leptonic,
arising from Compton-scattered ambient radiation.

\subsection{Microquasars and X-ray Binaries}

The discovery that LS 5039 is a TeV source \cite{aha05c} confirmed the
association of the EGRET source 3EG J1824-1514 with LS 5039
\cite{par00}. Its bolometric $\nu F_\nu$ flux in the 0.2 -- 10 TeV
range is at the level of $\approx 10^{-11}$ ergs cm$^{-2}$ s$^{-1}$.
The flux is modulated at the 3.9 day orbital period
\cite{aha06b}. Orbital modulation was also recently reported by the
MAGIC collaboration from the northern hemisphere microquasar LSI +61
303 \cite{alb06}, with a mean bolometric $\nu F_\nu$ flux of $\approx
4\times 10^{-11}$ ergs cm$^{-2}$ s$^{-1}$ in the 0.2 -- 4 TeV range
and mean photon spectral index $\alpha_{ph} \cong 2.5$.

Both of these sources are high-mass microquasars: the companion stars
in LS 5039 and LSI +61 303 have masses $\approx 23 M_\odot$ and
$\approx 10 M_\odot$, respectively.  The orbital modulation shows that
the TeV emission has to be produced in the vicinity of the binary
system.  Both leptonic \cite{brp06,db06} and hadronic models
\cite{aha06c} for microquasar jet emission have been proposed. Even
though their spectra are soft or cut off at multi-TeV energies, this
could be a result of $\gamma\gamma$ attenuation, so that the actual
$\gamma$-ray and neutrino production spectrum from hadronic
interactions could extend to very high energies \cite{aha06c,boe06}.

In addition to these two microquasars, the binary system B1259-63,
consisting of a pulsar and a high-mass Be star, is a TeV source
\cite{aha05d}. This raises the interesting possibility \cite{dub06}
that all three sources are binary systems consisting of a pulsar and a
high-mass star, with the TeV emission due to nonthermal particles
accelerated by the shock formed by interactions between the MHD wind
of the pulsar and the stellar wind of the high-mass star. The
formation of the emission in the inner region of the system would
naturally follow from this scenario.  Accreting X-ray binaries have
also been proposed \cite{anc06} as detectable neutrino sources when
protons and ions, accelerated in the magnetosphere of the system,
collide with material of the accretion disk, though the absence of TeV
emission suggests they would not be bright neutrino sources.

\subsection{Blazars}

The intense, highly variable $\gamma$-ray fluxes from blazars suggest
that they are also bright neutrino sources. It is important to
distinguish between the flat spectrum radio quasars (FSRQs), which
have strong atomic emission lines in their spectra, from the BL Lac
objects with their nearly featureless continua. All known TeV blazars
are X-ray selected BL Lac objects.

Though one might think that TeV blazars are the most probable neutrino
sources, given that particle acceleration to $\gg $ TeV energies must
take place in these sources, it is more likely that FSRQ blazars are
bright neutrino sources \cite{ad03}.  First, the brightest FSRQs have
$\nu F_\nu$ fluxes $\approx$ one order of magnitude brighter than the
TeV $\nu F_\nu$ fluxes of the brightest BL Lac objects. The absence of
TeV radiation in FSRQs is probably a consequence of $\gamma\gamma$
attenuation on the EBL, and does not indicate that TeV $\gamma$ rays
are not produced.  Moreover, the intense scattered accretion-disk
radiation in the vicinity of the supermassive black hole in FSRQs, as
indicated by the strong atomic lines, provides an important source of
target photons for photohadronic production \cite{ad01}.  Making the
conservative assumption that the energy injected in protons is equal
to the energy inferred from observations of the electron synchrotron
radio/X-ray emission, we  \cite{ad03,ad01} have shown that IceCube could detect one or
several neutrinos during bright FSRQ blazar flares, such as that
observed from 3C 279 in 1996.

\subsection{Gamma Ray Bursts}

This topic was recently reviewed \cite{da06}, and details can be found
there.  Neutrino production in GRBs depends most sensitively on two
parameters: the baryon-loading factor and Doppler factor of the GRB
blast wave. The baryon-loading factor refers to the energy in
nonthermal protons compared to the electromagnetic energy inferred
from direct measurements of the keV/MeV emission from GRBs. Provided
that the baryon-loading factor is $\gg 10$, which is required if GRBs
are the sources of UHECRs, and the Doppler factor is $\lesssim 200$,
neutrinos from GRBs are detectable with IceCube or a northern
hemisphere neutrino detector.  These calculations \cite{da03} are made
in the framework of the collapsar model, with values of the Doppler
factor in the range commonly expected for GRB outflows. Anomalous
$\gamma$-ray emission components \cite{hur94,gon03} in GRBs give
further evidence for hadronic acceleration by GRB blast waves.

\section{Diffuse Neutrinos}

Cosmic ray interactions in the disk of the Milky Way will make a
diffuse neutrino glow \cite{ber93}.  A flux of $\approx 160$ $ \nu$
$(>$ 250 TeV) km$^{-2}$ yr$^{-1}$ from a 5 square degree region
surrounding the galactic center is calculated in Ref.\ \cite{lm00},
though the exact result is sensitive to the hardness of the cosmic-ray
spectrum. Superpositions of emissions from the classes of point
sources listed above, in particular, GRBs and blazars, will make a
diffuse high-energy neutrino background radiation. Dim extragalactic
$\gamma$-ray sources can also make important neutrino backgrounds due
to their abundance.

Star-forming galaxies, which include normal spiral galaxies, starburst
galaxies, and infrared luminous galaxies, might be considered as
likely neutrino point sources because cosmic rays would certainly be
accelerated by the core-collapse supernovae resulting from the late
stages of evolution of the massive stars in these systems.  These
systems are, however, relatively dim $\gamma$-ray sources and,
furthermore, have soft spectra. The only extragalactic galaxy that was
detected with EGRET was the Large Magellanic Cloud \cite{sre92}, with
an integral $\gamma$-ray flux equal to $14.4(\pm 4.7) \times 10^{-8}$
ph($>100$ MeV cm$^{-2}$ s$^{-1}$ and a spectral index of $s = 2.2$,
implying a $\nu F_\nu$ flux of $\approx 2.3\times 10^{-11} (E/100$
MeV)$^{-0.2}$ ergs cm$^{-2}$ s$^{-1}$. Thus it would take $\gg 2$
years to detect neutrinos from the LMC unless there was an anomalous
hardening of the spectrum.

Nevertheless, the superpositions of the neutrino emissions from
star-forming galaxies will form a guaranteed background.  An estimate
\cite{lw06} of the neutrino background based on the synchrotron radio
luminosity associated with cosmic-ray acceleration in star-forming
galaxies is at a level detectable by IceCube, though the assumptions
and derived intensity have been challenged \cite{ste06}.

\begin{figure}[t]
\vskip-1.8in
\begin{minipage}{18pc}
\includegraphics[width=18pc]{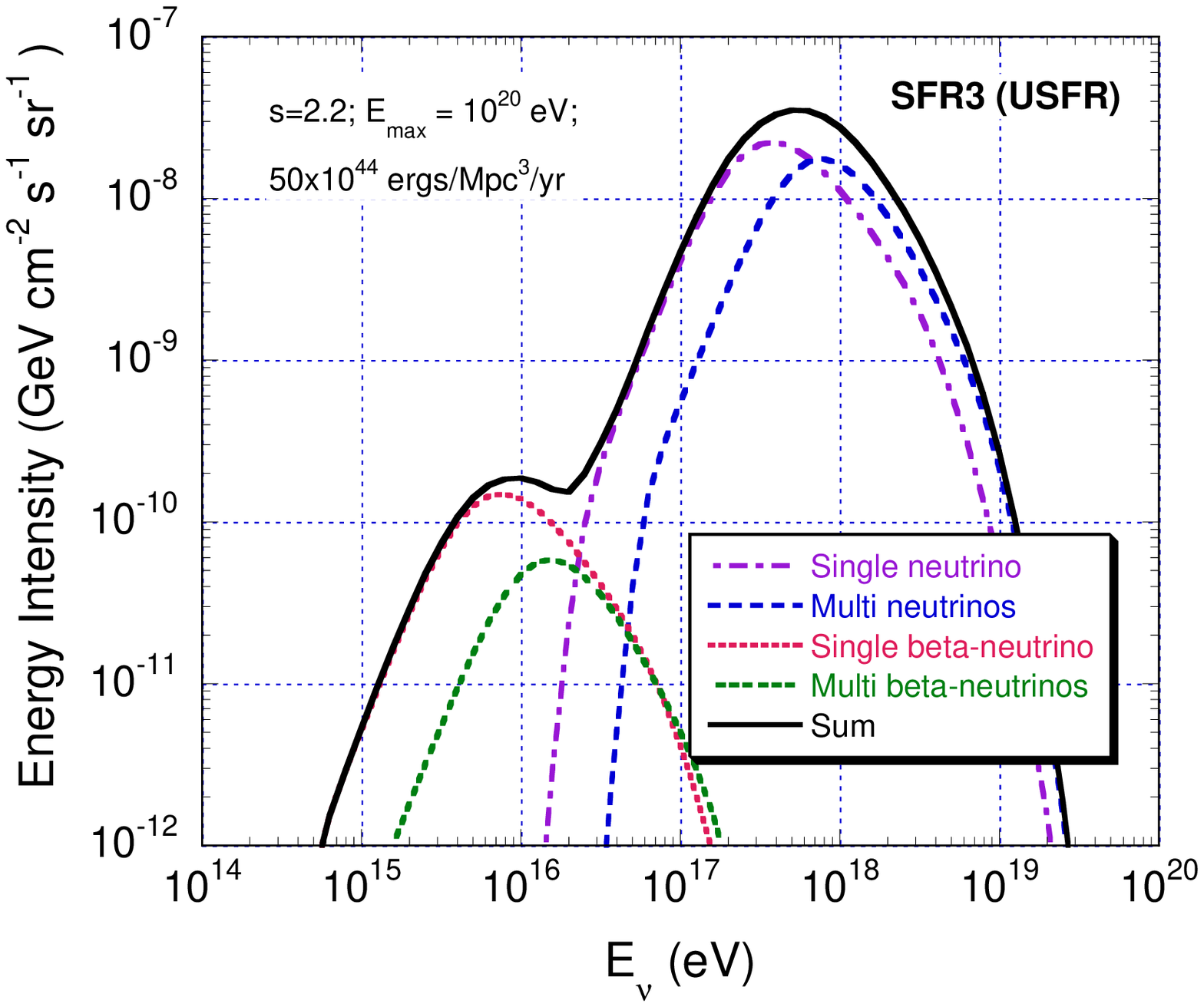}
\vskip-0.2in
\caption{\label{label}GZK neutrino flux using a star formation rate
history used to model the UHECR spectrum from GRBs \cite{wda04}.}
\end{minipage}\hspace{2pc}%
\begin{minipage}{18pc}
\includegraphics[width=18pc]{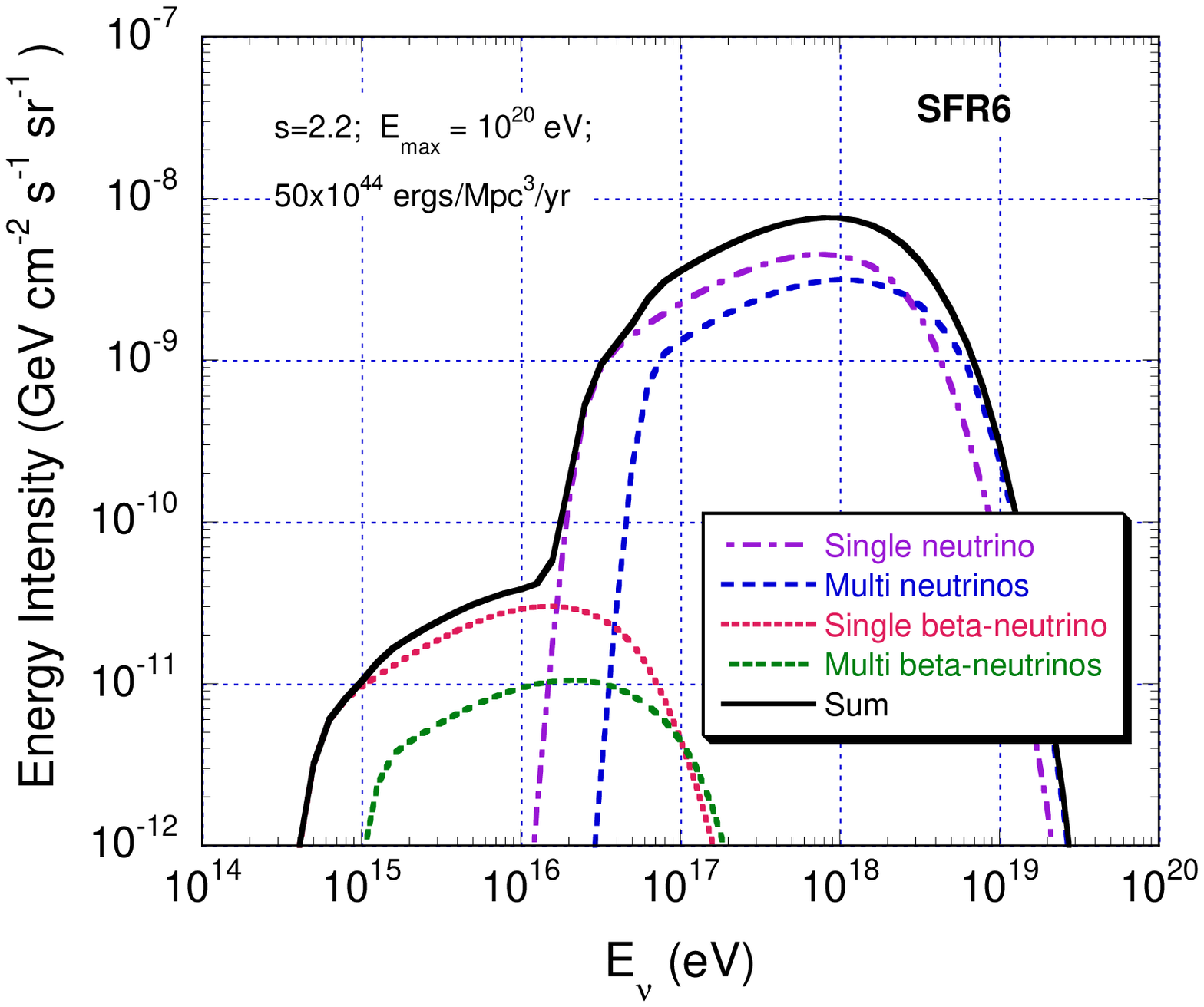}
\vskip-0.2in
\caption{\label{label}GZK neutrino flux using the star formation history
used to fit the statistics of GRBs with measured redshifts \cite{ld06}. }
\end{minipage} 
\end{figure}

Photohadronic interactions of UHECRs with photons of the EBL forms the
diffuse cosmogenic GZK neutrino flux. The spectrum of this background
depends on the UHECR activity in the early universe \cite{yk06}.
Figs.\ 1 and 2 show calculations (work in preparation with J.\ Holmes)
of the effect on the GZK neutrino background due to different
histories of GRB production.

\section{Summary}

The best-bet neutrino sources are GRBs, blazars, microquasars, and SNRs because of their
bright and hard $\gamma$-ray spectra that could originate from
hadronic processes. These results should be placed within the context
of theories of cosmic ray origin.  Supernova
remnants are thought to be the sources of cosmic rays to at least the
knee of the cosmic ray spectrum, because they have adequate power and
will produce strong shocks. But the TeV spectrum of RX J1713.7-3946
shows a cutoff at $\approx 10$ TeV.  Supernova remnants probably
differ greatly in their cosmic ray acceleration efficiencies, with
Type 1b/c supernovae and those associated with GRBs being the
strongest such accelerators.  Microquasars and Be/X-ray binaries,
though less powerful than SNRs, will also form strong shocks and will
have stellar wind material that can provide a target for secondary
nuclear production. 

UHECRs must be extragalactic given the $\sim \mu$G strength of the
Milky Way's magnetic field, so that GRBs and blazars, the brightest
extragalactic $\gamma$-ray sources, are the most probable sources of
UHECRs and high-energy neutrinos. The highly relativistic outflows and
shocks required to model these systems can accelerate particles to
$\approx 10^{20}$ eV.

\ack This work is supported by the Office of Naval Research. 

\end{document}